\documentclass{article}

\usepackage{amsmath}
\usepackage{amssymb}
\usepackage{amsthm}
\usepackage{tikz-cd}
\usepackage{ifthen}
\usepackage{xspace}
\usepackage{xparse}
\usepackage[pdftex,
            pdfauthor={Jason Z. S. Hu},
            pdftitle={Internal Category with Families in Presheaves}]{hyperref}
\usepackage{mathpartir}
\usepackage{cleveref}

\allowdisplaybreaks

\theoremstyle{definition}
\newtheorem{definition}{Definition}[section]

\newcommand{\ctxext}[1]{\langle #1 \rangle}
\newcommand{\To}{\Rightarrow}
\newcommand{\D}{\ensuremath{\mathcal{D}}\xspace}
\newcommand{\Set}{\ensuremath{\texttt{Set}\xspace}}
\newcommand{\Nat}{\ensuremath{\texttt{Nat}}}
\newcommand{\THom}{\ensuremath{\texttt{Hom}}}
\newcommand{\TApp}{\ensuremath{\texttt{App}}}
\newcommand{\tbox}{\ensuremath{\texttt{box}}}
\newcommand{\tletbox}{\ensuremath{\texttt{letbox}}}
\newcommand{\Ctx}{\ensuremath{\texttt{Ctx}}}
\newcommand{\TTy}{\texttt{Ty}}
\DeclareDocumentCommand{\STy}{o}{
  \IfNoValueTF{#1}{
    \TTy
  } {
    \TTy(#1)
  }
}
\newcommand{\TTm}{\texttt{Tm}}
\DeclareDocumentCommand{\STm}{o o}{
  \IfNoValueTF{#1}{
    \TTm
  } {
    \TTm(#1, #2)
  }
}

\DeclareDocumentCommand{\hSTy}{o}{
  \IfNoValueTF{#1}{
    \widehat\TTy
  } {
    \widehat\TTy(#1)
  }
}
\DeclareDocumentCommand{\hSTm}{o o}{
  \IfNoValueTF{#1}{
    \widehat\TTm
  } {
    \widehat\TTm(#1, #2)
  }
}

\newcommand{\UHom}{\ensuremath{\textsf{Hom}}}
\newcommand{\UApp}{\ensuremath{\textsf{App}}}
\newcommand{\ubox}{\ensuremath{\textsf{box }}}
\newcommand{\uletbox}[3]{\ensuremath{\textsf{let box }#1 = #2\textsf{ in }#3}}
\newcommand{\UTy}{\textsf{Ty}}
\DeclareDocumentCommand{\VTy}{o}{
  \IfNoValueTF{#1}{
    \UTy
  } {
    \UTy(#1)
  }
}
\newcommand{\UTm}{\textsf{Tm}}
\DeclareDocumentCommand{\VTm}{o o}{
  \IfNoValueTF{#1}{
    \UTm
  } {
    \UTm(#1, #2)
  }
}

\DeclareDocumentCommand{\judge}{ o m } {
  \IfNoValueTF {#1}
  {\Gamma \vdash #2}
  {#1 \vdash #2}
}

\DeclareDocumentCommand{\mjudge}{ o o m } {
  \IfNoValueTF {#1}
  {\IfNoValueTF {#2}
    {\Delta ; \Gamma \vdash #3}
    {\Delta; #2 \vdash #3}}
  {\IfNoValueTF {#2}
    {#1 ; \Gamma \vdash #3}
    {#1; #2 \vdash #3}}
}

\DeclareDocumentCommand{\typing}{ o m m } {
  \judge[#1]{#2 : #3}
}

\DeclareDocumentCommand{\istype}{o m} {
  \judge[#1]{#2\texttt{ type}}
}

\DeclareDocumentCommand{\ismtype}{o o m} {
  \mjudge[#1][#2]{#3\texttt{ type}}
}

\DeclareDocumentCommand{\mtyping}{ o o m m } {
  \mjudge[#1][#2]{#3 : #4}
}

\newcommand{\whhat}[1]{\widehat{#1}}
\newcommand{\wtop}{\whhat\top}

\title{Internal Category with Families in Presheaves}
\date{}
\author{Jason Z. S. Hu \\ McGill University \\ \href{mailto:zhong.s.hu@mail.mcgill.ca}{zhong.s.hu@mail.mcgill.ca}}

\begin{document}
\maketitle

\begin{abstract}
  In this note, we review a construction of category with families (CwF) in a presheaf
  category. When the base category of a presheaf category is a CwF, we internalize
  this CwF structure in the CwF of the presheaf category. This note assumes working
  knowledge on category theory.
\end{abstract}

\section{Category with Families}

Category with families (CwF)~\cite{DBLP:conf/types/Dybjer95} is a categorical
structure to model dependent type theories. Essentially it characterizes substitution
invariance of dependent type theories in a categorical language.

\begin{definition}
  A category \D is a category with families if it has
  \begin{enumerate}
  \item a terminal object $\top$ and $!$ is the unique morphism to $\top$,
  \item a functor $\STy : \D^{op} \To \Set$; given $\Gamma, \Delta : \D, A :
    \STy[\Gamma], \sigma : \Delta \To \Gamma$, then we write the functorial action of $\STy$
    as $A\{\sigma\} : \STy[\Delta]$,
  \item given $\Gamma : \D, A : \STy[\Gamma]$, a set $\STm[\Gamma][A]$; given
    $\Gamma, \Delta : \D, A : \STy[\Gamma], \sigma : \Delta \To \Gamma, M :
    \STm[\Gamma][A]$, then $M\{\sigma\} : \STm[\Delta][A\{\sigma\}]$,
  \item given $\Gamma : \D, A : \STy[\Gamma]$, their context comprehension $\Gamma.A
    \in D$, 
  \item given $\Gamma : \D, A : \STy[\Gamma]$, the projection morphism $p_A : \Gamma.A
    \To \Gamma$,
  \item given $\Gamma : \D, A : \STy[\Gamma]$, the variable projection $v_A :
    \STm[\Gamma.A][A\{p_A\}]$, and
  \item given $\Gamma, \Delta : \D, A : \STy[\Gamma], \sigma : \Delta \To \Gamma, M :
    \STm[\Delta][A\{\sigma\}]$, then a substitution extension $\ctxext{\sigma, M} :
    \Delta \To \Gamma.A$,
  \end{enumerate}
  such that the following equations hold
  \begin{enumerate}
  \item $p_A \circ \ctxext{\sigma, M} = \sigma$,
  \item $v_A\{\ctxext{\sigma, M}\} = M$, and
  \item for $\sigma : \Delta \To \Gamma.A$, $\sigma = \ctxext{p_A \circ \sigma, v_A\{\sigma\}}$.
  \end{enumerate}
\end{definition}

Intuitively, CwFs are ``substitution-oriented''. Objects in $\D$ model contexts in
type theories and morphisms model substitutions between the contexts. Hence we need
the terminal object to model the empty context. $\STy[\Gamma]$ represents the set of
semantic types in context $\Gamma$ and given $A : \STy[\Gamma]$, $\STm[\Gamma][A]$ is
the set of semantic terms of type $A$ in $\Gamma$. Comprehensions extend contexts. The
operations like projections and extensions are used to manipulate substitutions. $v$
is sometimes called the second projection, which models
variables. 

\section{Presheaves as CwFs}\label{sec:presh}

\cite{Hofmann1997} showed that any given presheaf category has a CwF structure. Here
we assume any given base category $\D$ and specify the construction in details.

Conventionally, we refer to the presheaf category as $\whhat\D$. Moreover, we put
iterative applications in a single pair of parentheses, namely
\begin{align*}
  f(x_1, \cdots, x_n) := f(x_1)\cdots(x_n)
\end{align*}

\subsection{Terminal Object}

All presheaf category has a terminal object. This is defined via
\begin{align*}
  \top &: \D^{op} \To \Set \\
  \top(d : \D) &:= \{*\} \\
  \top(\delta : \D(d, d')) &:= * \mapsto * : \{ * \} \to \{ * \}
\end{align*}

The unique morphism into $\top$ is a natural transformation:
\begin{align*}
  ! &: \Nat(\Gamma, \top) \\
  !(d : \D) &:= s \mapsto * : \Gamma(d) \to \{ * \}
\end{align*}

Since the codomain is $\{*\}$, we know that $!$ must be unique.

\subsection{Types}

Semantic types is modelled by the presheaf functor $\STy : \whhat\D \To \Set$. Here
the intension is to define the codomain as the set presheaf over the category of
elements, as specified below.

\begin{definition}
  Given $\Gamma : \whhat\D$, the category of elements $\smallint\Gamma$ has the following
  data:
  \begin{enumerate}
  \item objects are $\Sigma (d : \D) \Gamma(d)$,
  \item morphisms between $(d : \D , s : \Gamma(d)) \To (d' : \D, s' : \Gamma(d'))$ is
    a morphism $\delta : d \To d'$, such that $\Gamma(\delta, s') = s$. Set
    theoretically, we write
    \begin{align*}
      (d : \D , s : \Gamma(d)) \To (d' : \D, s' : \Gamma(d')) := \{ \delta : d \To d'
      | \Gamma(\delta, s') = s\}
    \end{align*}
  \item identities and compositions are inherited from $\D$. The well-definedness of
    compositions follows from the functoriality of $\Gamma$. 
  \end{enumerate}

  Since identities and compositions are inherited from $\D$, the categorical laws are
  also inherited from $\D$. 
\end{definition}
Here we employ some type theoretical notions to ease the discussion. We can easily
port the language to a more set theoretical one, if necessary. 

Notice that, given $s' : \Gamma(d')$, we can extend $\delta : \D(d, d')$ to a morphism
in $\smallint \Gamma$: $\delta : (d, \Gamma(\sigma, s')) \To (d', s')$. This observation will
come in handy later. 

We then move on to define $\STy$:
\begin{align*}
  \STy &: \whhat\D \To \Set \\
  \STy(\Gamma) &:= (\smallint\Gamma)^{op} \To \Set \\ 
  \_\{\_\} &: \STy(\Gamma) \to \Delta \To \Gamma \to \STy(\Delta) \\
  A\{\sigma\}(d : \D, s : \Delta(d)) &:= A(d, \sigma(d, s)) \\
  A\{\sigma\}(\delta : (d, s) \To (d', s')) &:= A(\delta) : A\{\sigma\}(d, s) \to
                                              A\{\sigma\}(d', s')
\end{align*}
As mentioned in the definition of CwFs, we write $A\{\sigma\}$ for  $\STy(\sigma, A)$ so we have given
both components of $\STy$. Now we need to verify the well-definedness of this
definition. First, we need to examine $A\{\sigma\}(\delta)$ is well defined. It is 
well-defined only when $u$ and $u'$ in $\delta : (d, u : \Gamma(d)) \To (d', u' :
\Gamma(d'))$ are clearly specified on the right hand side. This can be done by requiring
\begin{align*}
  u &:= \sigma(d, s) \\
  u' &:= \sigma(d', s')
\end{align*}
which also aligns with the object part of $A\{\sigma\}$. We also need to prove
$\Gamma(\delta, u') = u$ knowing $\Delta(\delta, s') = s$.  We analyze as follows:
\begin{align*}
  \Gamma(\delta, u')
  &= \Gamma(\delta, \sigma(d', s')) \\
  &= \sigma(d, \Delta(\delta, s')) \tag*{by naturality of $\sigma$} \\
  &= \sigma(d, s) \tag*{by the equation above} \\
  &= u
\end{align*}
That concludes that $A\{\sigma\}$ is well defined.

Given both components of $\STy$, we then move on to examine the functorial laws of
$\STy$. Since the morphism part of $A\{\sigma\}$ directly inherits from $A$, we only
need to examine the object part. Consider the identity law
\begin{align*}
  A\{id\}(d, s) &= A(d, id(d, s)) = A(d, s)
\end{align*}

Next we examine the composition law:
\begin{align*}
  A\{\sigma \circ \sigma'\}(d, s) &= A(d, (\sigma \circ \sigma')(d, s)) \\
                                  &= A(d, \sigma(d, \sigma'(d, s))) \\
  A\{\sigma\}\{\sigma'\}(d, s) &= A\{\sigma\}(d, \sigma'(d, s)) \\
                                  &= A(d, \sigma(d, \sigma'(d, s)))
\end{align*}
Thus they agree. At this point, we can conclude $\STy$ is a presheaf.

\subsection{Terms}

Semantic terms are dependent objects which are indexed semantic contexts and semantic
types. We define the set of terms as follows:
\begin{align*}
  \STm[\Gamma][A] := \{& M : (o : \Sigma (d : \D) \Gamma(d)) \to A(o) \\
  &| \forall \delta
  : d' \To d, s : \Gamma(d). A(\delta, M(d, s)) = M(d', \Gamma(\delta, s)) \}
\end{align*}

Now consider term substitutions. As typically done, we overload the notation for type
substitutions:
\begin{align*}
    \_\{\_\} &: \STm[\Gamma][A] \to (\sigma : \Delta \To \Gamma) \to
               \STm[\Delta][A\{\sigma\}] \\
  M\{\sigma\}(d, s : \Delta(d)) &:= M(d, \sigma(d, s) : \Gamma(d)) 
\end{align*}
There are two things to verify for this definition. We need to show that the right
hand side is in the set $A\{\sigma\}(d, s)$ and that the set specification holds. The first
item holds by definition. Now we examine the set specification.

The set specification requires that given $\delta : d' \To d$ and $s : \Delta(d)$, \linebreak
$A\{\sigma\}(\delta, M\{\sigma\}(d, s)) = M\{\sigma\}(d', \Delta(\delta, s))$ holds. Let us
consider the left hand side:
\begin{align*}
  A\{\sigma\}(\delta, M\{\sigma\}(d, s)) &= A\{\sigma\}(\delta, M(d, \sigma(d, s)))
                                           \tag*{by definition of $M\{\sigma\}$} \\
                                         &= A(\delta, M(d, \sigma(d, s))) \tag*{by
                                           definition of $A\{\sigma\}$} \\
                                         &= M(d', \Gamma(\delta, \sigma(d, s)))
                                           \tag*{by $M$'s specification}
\end{align*}
Now we analyze the right hand side:
\begin{align*}
  M\{\sigma\}(d', \Delta(\delta, s)) &= M(d', \sigma(d', \Delta(\delta, s))) \tag*{by definition}
\end{align*}
Both sides would agree if $\Gamma(\delta, \sigma(d, s)) = \sigma(d', \Delta(\delta,
s))$, and this holds due to the naturality of $\sigma$. 

This concludes that $M\{\sigma\}$ is a well-defined semantic term.

\subsection{Context Comprehension}

Context comprehension models extension of context in the syntax. The operator $.$
joins a context $\Gamma$ and a type $A$, forming another context $\Gamma.A$. Given
$\Gamma : \whhat\D$ and $A : \STy[\Gamma]$, we need to give $\Gamma.A : \whhat\D =
\D^{op} \To \Set$.
\begin{align*}
  \Gamma.A &: \whhat\D \\
  \Gamma.A(d \in \D^{op}) &:= \Sigma (s : \Gamma(d)) A(d, s) \\
  \Gamma.A(\delta : \D^{op}(d, d'))(s : \Gamma(d), a : A(d, s))
           &:= (\Gamma(\delta, s), A(\delta, a)) : \Sigma(s' : \Gamma(d')) A(d', s')
\end{align*}

Now we examine the well-definedness of this presheaf. We need to ensure that the
morphism part is well defined, in particular the second component of the result indeed
resides in $A(d', s') = A(d', \Gamma(\delta, s))$. 

Notice that $\delta : D^{op}(d, d') = D(d', d)$ and we are given $s :
\Gamma(d)$. Recall in the previous subsection, we discussed that $\delta$ can be
extended to a morphism in $\smallint \Gamma$, so we know
$\delta : \smallint\Gamma((d', \Gamma(\delta, s)), (d, s))$. Since $A$ is a presheaf
over $\smallint\Gamma$, then $A(\delta)$ is a set function
$A(d, s) \to A(d', \Gamma(\delta, s))$. This justifies that $A(\delta, a)$ does reside
in $A(d', \Gamma(\delta, s))$.

Now we examine the functorial laws. This is immediate because $\Gamma.A$ simply
combines $\Gamma$ and $A$, and thus the functoriality is induced from the
functoriality of $\Gamma$ and $A$.

\subsection{Projection}

Given $\Gamma$ and $A : \STy[\Gamma]$, the projection morphism
$p_A : \Gamma.A \To \Gamma$ models the weakening substitution. Since it is a morphism
between two presheaves, it is a natural transformation. Its definition is easily given
due to the definition of $\Gamma.A$:
\begin{align*}
  p_A &: \Gamma.A \To \Gamma \\
  p_A(d)(s : \Gamma(d), \_) &:= s
\end{align*}

This definition is obviously natural.

\subsection{Variables}

Given $\Gamma$ and $A : \STy[\Gamma]$, $v_A : \STm[\Gamma.A][A\{p_A\}]$ models the
first variable in the context. Combining the projection morphism, we can model any
variables. In the presheaf category, we can model this term as:
\begin{align*}
  v_A &: \STm[\Gamma.A][A\{p_A\}] \\
  v_A(d : \D, (s : \Gamma(d), a : A(d, s))) &:= a
\end{align*}
Now we need to show that $a : A\{p_A\}(d, (s, a))$. We expand the set by definition:
\begin{align*}
  A\{p_A\}(d, (s, a)) &= A(d, p_A(s, a)) = A(d, s)
\end{align*}

Next we shall examine that this definition complies with the set specification given
by $\STm$. Assuming $\delta : d' \To d$, $s : \Gamma(d)$ and $a : A(d, s)$, we need that
$A\{p_A\}(\delta, v_A(d, (s, a))) = v_a(d', \Gamma.A(\delta, (s, a)))$. We analyze
each side:
\begin{align*}
  A\{p_A\}(\delta, v_A(d, (s, a)))
  &= A\{p_A\}(\delta, a) \tag*{by definition of $v_A$} \\
  &= A(\delta, a) \tag*{by definition of $A\{p_A\}$}
\end{align*}
\begin{align*}
  v_a(d', \Gamma.A(\delta, (s, a)))
  &= v_a(d', (\Gamma(\delta, s), A(\delta, a))) \tag*{by definition of $\Gamma.A$} \\
  &= A(\delta, a) \tag*{by definition of $v_A$} \\
\end{align*}
Thus the specification is satisfied.

\subsection{Substitution Extension}

The last piece of data is the substitution extension. Given $\sigma : \Delta \To
\Gamma$, $A : \STy[\Gamma]$ and $M : \STm[\Delta][A\{\sigma\}]$, we want to obtain
$\ctxext{\sigma, M} : \Delta \To \Gamma.A$.
\begin{align*}
  \ctxext{\sigma, M} &: \Delta \To \Gamma.A \\
  \ctxext{\sigma, M}(d : \D, s : \Delta(s)) &:= (\sigma(d, s), M(d, s)) : \Gamma.A(d) = \Sigma (s' :
                                              \Gamma(d))A(d, s')
\end{align*}

Next we shall see that $\ctxext{\sigma, M}$ is natural. This is to examine that given
$\delta : \D^{op}(d, d')$ and $s : \Delta(d)$, we have the equation $\ctxext{\sigma,
  M}(d', \Delta(\delta, s)) = \Gamma.A(\delta, \ctxext{\sigma, M}(d, s))$. We analyze
both sides:
\begin{align*}
  \ctxext{\sigma, M}(d', \Delta(\delta, s))
  &= (\sigma(d', \Delta(\delta, s)), M(d', \Delta(\delta, s))) \tag*{by definition of
    $\ctxext{\sigma, M}$}
\end{align*}
\begin{align*}
  \Gamma.A(\delta, \ctxext{\sigma, M}(d, s))
  &= \Gamma.A(\delta, (\sigma(d, s), M(d, s))) \tag*{by definition of
    $\ctxext{\sigma, M}$} \\
  &= (\Gamma(\delta, \sigma(d, s)), A(\delta, M(d, s))) \tag*{by definition of
    $\Gamma.A$} \\
  &= (\sigma(d', \Delta(\sigma, s)), A(\delta, M(d, s))) \tag*{by naturality of
    $\sigma$} \\
  &= (\sigma(d', \Delta(\delta, s)), M(d', \Delta(\delta, s))) \tag*{by specification
    of $M$}
\end{align*}
Thus $\ctxext{\sigma, M}$ is natural.

\subsection{Laws}

The definition of CwFs also requires laws to hold. Next we examine each of them.  In
the following discussion, we assume $\sigma : \Delta \To \Gamma$, $A : \STy[\Gamma]$
and $M : \STm[\Delta][A\{\sigma\}]$.
\begin{align*}
  (p_A \circ \ctxext{\sigma, M})(d, s)
  = p_A (\ctxext{\sigma, M}(d, s)) = \sigma(d, s)
\end{align*}
Thus $p_A \circ \ctxext{\sigma, M} = \sigma$.
\begin{align*}
  v_A\{\ctxext{\sigma, M}\}(d, s)
  &= v_A(d, \ctxext{\sigma, M}(d, s)) \tag*{by definition of term substitution} \\
  &= M(d, s) \tag*{by definition of $\ctxext{\sigma, M}$ and $v_A$}
\end{align*}
Thus $v_A\{\ctxext{\sigma, M}\} = M$.

Given $\sigma' : \Delta \To \Gamma.A$, we have
\begin{align*}
  &\ \ctxext{p_A \circ \sigma', v_A\{\sigma'\}}(d, s) \\
  =&\ ((p_A \circ \sigma')(d, s), v_A\{\sigma'\}(d, s)) \tag*{by definition of
    substitution extension}\\
  =&\ (\pi_1 (\sigma'(d, s)), v_A\{\sigma'\}(d, s))
    \tag*{$\pi_1$ is the first projection of a $\Sigma$ set} \\
  =&\ (\pi_1 (\sigma'(d, s)), \pi_2(\sigma'(d, s)))
    \tag*{by definition of $v_A\{\sigma'\}$ and $\pi_2$ is the second projection of a
    $\Sigma$} \\
  =&\ \sigma'(d, s) \tag*{extensionality of $\Sigma$}
\end{align*}
This gives $\ctxext{p_A \circ \sigma', v_A\{\sigma'\}} = \sigma'$. 

\section{Internal CwFs}\label{sec:internal}

Let us consider the situation where the (small) base category $\D$ is a CwF. To avoid
confusions with notations, we employ the following conventions:
\begin{table}[h]
\begin{tabular}{llllll}
                              & Contexts         & Set of Types    & Types     & Set of Terms    & Terms     \\\hline
Base $\D$                     & $\Phi, \Psi$     & $\STy$          & $S, T, U$ & $\STm$          & $s, t, u$ \\
Presheaves $\whhat \D$      & $\Gamma, \Delta$ & $\whhat \STy$ & $A, B$    & $\whhat \STm$ & $M, N$    \\
Internal CwF in $\whhat \D$ & $\Phi, \Psi$     & $\VTy$          & $S, T, U$ & $\VTm$          & $s, t, u$
\end{tabular}
\end{table}

Before entering the discussion, we would like to fix the terminologies. There are
types and terms referring to those in three different contexts. We use semantic types
(resp. terms) for the types (resp. terms) in $\whhat\D$. We refer to types
(resp. terms) in the internal CwF as internal types (resp. terms). When referring to
those in $\D$, we will explicitly call them types (resp. terms) in $\D$. 

Our purpose in this section is to \emph{internalize} the CwF structure of $\D$ in the
presheaf category $\whhat\D$. We call this internal structure the internal CwF. Now
in the categorical part, there are two CwF structures: the one of $\D$ and the one of
$\whhat \D$ specified in the previous section. Therefore, following the convention,
we use $\STy$ and $\STm$ to represent the semantic types and terms in $\D$,
respectively, and we put hats over and use $\hSTy$ and $\hSTm$ for semantic types and
terms in $\whhat\D$, respectively. In the later part of the section, we need to
define the internalized version of CwF of $\D$, the types and terms of which are
represented by $\hSTy$ and $\hSTm$.

To be more specific, by internalizing the CwF of $\D$, we mean to be able to study the
CwF inside of $\whhat\D$. That requires to represent types and terms of $\D$ in
$\whhat\D$. Judgmentally, we should define semantic types and terms which can suitably
interpret the following judgments:
\begin{mathpar}
  \inferrule*
  { }
  {\istype[]{\Ctx}}

  \inferrule*
  { }
  {\istype[\Psi : \Ctx, \Phi : \Ctx]{\UHom(\Psi, \Phi)}}

  \inferrule*
  { }
  {\istype[\Psi : \Ctx]{\VTy[\Psi]}}

  \inferrule*
  { }
  {\istype[\Psi : \Ctx, A : \VTy[\Psi]]{\VTm[\Psi][A]}}
\end{mathpar}

Now we construct suitable semantic types in $\whhat \D$ for interpreting these
syntactic types.

\subsection{Internal Contexts}

Now we define the internalized representation for objects in $\D$.
\begin{align*}
  \Ctx &: \hSTy[\wtop] \\
  \Ctx(\_) &:= Obj(\D) \\
  \Ctx(\delta) &:= \Psi \mapsto \Psi
\end{align*}

To actually show that this definition does represent the objects of $\D$, we need to
show that $\hSTm[\wtop][\Ctx] \simeq Obj(\D)$. 

By definition,
\begin{align*}
  \hSTm[\wtop][\Ctx] = \{& M : (o : \Sigma (\Psi : \D) \wtop(\Psi)) \to \Ctx(o) \\
  & | \forall \delta : \Phi \To \Psi, s : \wtop(\Psi). \Ctx(\delta, M(\Psi, s)) = M(\Phi, \wtop(\delta,
  s)) \}
\end{align*}
After simplification, we have
\begin{align*}
  \hSTm[\wtop][\Ctx] = \{& M : (o : \Sigma (\Psi : \D) \{*\}) \to Obj(\D) \\
  &  \forall \delta : \Phi \To \Psi, s : \{*\}. M(\Psi, s) = M(\Phi, s) \}
\end{align*}

That is, $M$ is invariant under different $\Psi$. Since we know $\D$ is a CwF, we know it
must have a terminal object $\top$. Now we can show the intended isomorphism:
\begin{align*}
  f &: \hSTm[\wtop][\Ctx] \to Obj(\D) \\
  f(M) &:= M(\top, *) \\
  g &: Obj(\D) \to \hSTm[\wtop][\Ctx] \\
  g(\Psi)(\_) &:= \Psi
\end{align*}
Clearly, $f$ and $g$ do form an isomorphism and thus this representation of $\Ctx$ is
as intended.

We can generalize the construction in this subsection. Reviewing the reason why
$\hSTm[\wtop][\Ctx] \simeq Obj(\D)$, we realize that if a $\hSTy$ is defined
without considering the first projection in the object part, then the $\hSTm$ is
automatically isomorphic to the specified set given by $\hSTy$. Moreover, the morphism
part then must be identity function. We will also see this pattern in the following subsections.

\subsection{Internal Hom}

Since $\D$ is a CwF, the morphisms in it are substitutions. We can also model this in
$\whhat\D$.
\begin{align*}
  \UHom &: \hSTy[\wtop . \Ctx . \Ctx\{p_\Ctx\}] \\
  \UHom(\_, (*, \Psi, \Phi)) &:= \THom(\Psi, \Phi) \\
  \UHom(\delta) &:= \delta' \mapsto \delta'
\end{align*}
Here we have two notions of Homs and we distinguish them by fonts. $\THom$ represents
the Hom-set in $\D$ while $\UHom$ is its internalization. Again, here we will make use
of the observation at the end of the last subsection. Since the definition of the
object part of $\UHom$ is independent of the first projection, we know that its
$\hSTm$ is automatically isomorphic to $\THom(\Psi, \Phi)$. We can verify this
observation more formally.
\begin{align*}
  \hSTm[\wtop.\Ctx.\Ctx\{p_\Ctx\}][\UHom] =
  \{& M : (o : \Sigma (\Psi : \D)(\wtop.\Ctx.\Ctx\{p_\Ctx\}(\Psi))) \to \UHom(o) \\
  | & \forall \delta : \Phi \To \Psi, s : \wtop.\Ctx.\Ctx\{p_\Ctx\}(\Psi). \\
    & \UHom(\delta, M(\Psi, s)) = M(\Phi, \wtop.\Ctx.\Ctx\{p_\Ctx\}(\delta, s))  \}
\end{align*}
Let us analyze the specification, the left hand side equals to $M(\Psi, s)$. By definition
of context comprehension, we know $s = (*, \Psi', \Phi')$ for some $\Psi'$ and
$\Phi'$. The right hand side becomes
\begin{align*}
  &\ M(\Phi, \wtop.\Ctx.\Ctx\{p_\Ctx\}(\delta, s)) \\
  =&\ M(\Phi, \wtop.\Ctx.\Ctx\{p_\Ctx\}(\delta, (*, \Psi', \Phi')))
    \tag*{extensionality for $s$} \\
  =&\ M(\Phi, (*, \Psi', \Phi'))
    \tag*{functorial actions of $\wtop$ and $\Ctx$ are identity} \\
  =&\ M(\Phi, s)
\end{align*}
That is, again, $M$ is invariant under different $\Psi$. That allows us to conclude
$\hSTm[\wtop.\Ctx.\Ctx\{p_\Ctx\}][\UHom] \simeq (\Psi, \Phi : \D) \to \THom(\Psi,
\Phi)$.

\subsection{Internal Category}

Before moving on, we shall see that the whole categorical structure of $\D$ can be
internalized in $\whhat\D$. That is to say we should find semantic terms that can interpret the
following judgments:
\begin{mathpar}
  \inferrule*
  { }
  {\typing[\Psi : \Ctx]{id_\Psi}{\UHom(\Psi, \Psi)}}
  
  \inferrule*
  { }
  {\typing[\Psi : \Ctx, \Psi' : \Ctx, \Psi'', \Ctx, \sigma' : \UHom(\Psi', \Psi''),
    \sigma : \UHom(\Psi, \Psi')]{\sigma' \circ \sigma}{\UHom(\Psi, \Psi'')}}
\end{mathpar}

These two terms correspond to the following terms (after expanding the definitions of
functorial actions):
\begin{align*}
  id &: \Sigma (\Phi : \D) (\Sigma (s : \{*\}) (\Psi : \D)) \to \THom(\Psi, \Psi) \\
  id(\_, (*, \Psi)) &:= id_\Psi
\end{align*}
\begin{align*}
  \circ &: \Sigma (\Phi : \D) (\Sigma (\Sigma (\Sigma (\Sigma (\Sigma (s : \{*\}) (\Psi : \D)) (\Psi'
          : \D)) (\Psi'' : \D)) \\
        &\ \ (\sigma' : \THom(\Psi', \Psi''))) (\sigma : \THom(\Psi, \Psi'))) \\
        &\ \ \to \THom(\Psi, \Psi'') \\
  \circ(\_, (*, \Psi, \Psi', \Psi'', \sigma', \sigma)) &:= \sigma' \circ \sigma
\end{align*}
In above definitions, $id_\Psi$ and $\sigma' \circ \sigma$ on the right hand side are
in $\D$. Based on this definition, we can see that the categorical laws hold
internally as well. These definitions are valid semantic terms because they do not
depend on the first projection, and by the analysis above this already implies the
validity of the definitions as terms. In later constructions of semantic terms in this
section, we always ignore the first projection and thus they are always valid. For
this reason, we will omit the same explanation. 

\subsection{Internal Types}

Next we internalize $\STy$ to $\whhat\D$, which is represented by $\VTy$.
\begin{align*}
  \VTy &: \hSTy[\wtop.\Ctx] \\
  \VTy(\_, (*, \Psi)) &:= \STy(\Psi) \\
  \VTy(\delta) &:= S \mapsto S
\end{align*}

Similarly, we should also examine $\hSTm[\wtop.\Ctx][\VTy]$. By definition
\begin{align*}
  \hSTm[\wtop.\Ctx][\VTy]
  = \{& M : (o : \Sigma (\Psi : \D) \wtop.\Ctx(\Psi)) \to \VTy(o) \\
      & | \forall \delta : \Phi \To \Psi, s : \wtop.\Ctx(\Psi).
        \VTy(\delta, M(\Psi, s)) = M(\Phi, \wtop.\Ctx(\delta, s))\}
\end{align*}
Similarly, we can show that $M$ is invariant under different $\Psi$ and thus we obtain
the isomorphism
$\hSTm[\wtop.\Ctx][\VTy] \simeq (\Psi : \D) \to \STy(\Psi)$.

Next, we should show that substitution can also be internalized. We are looking to
internalize the following judgment:
\begin{align*}
  \inferrule*
  { }
  {\typing[\Psi : \Ctx, \Phi : \Ctx, T : \VTy[\Phi], \sigma : \UHom(\Psi, \Phi)]{T\{\sigma\}}{\VTy[\Psi]}}
\end{align*}
This correspond to defining the following function $M$: 
\begin{align*}
  M &: \Sigma (\Psi' : \D) (\Sigma (\Sigma (\Sigma (\Sigma (s : \{ *\}) (\Psi : \D)) (\Phi :
      \D)) \\
    &\ \ (T : \STy(\Phi))) \THom(\Psi, \Phi))
      \to \STy(\Psi) \\
  M(\_, (* , \Psi, \Phi, T, \sigma)) &:= T\{\sigma\}
\end{align*}
The substitution on the right hand side is given by the type substitution of CwF of
$\D$. 

\subsection{Internal Terms}

In the same spirit, we can internalize $\STm$ to $\whhat\D$:
\begin{align*}
  \VTm &: \hSTy[\wtop.\Ctx.\VTy] \\
  \VTm(\_, (*, \Psi, S)) &:= \STm(\Psi, S) \\
  \VTm(\delta) &:= s \mapsto s
\end{align*}
Moreover, we can also show $\hSTm[\wtop.\Ctx.\VTy][\VTm] \simeq (\Psi : \D)(S : \STy(\Psi)) \to
\STm(\Psi, S)$. 

We can also formulate term substitutions. Our target is the following judgment:
\begin{align*}
  \inferrule*
  { }
  {\typing[\Psi : \Ctx, \Phi : \Ctx, T : \VTy[\Phi], t : \VTm[\Phi][T], \sigma :
  \UHom(\Psi, \Phi)]{t\{\sigma\}}{\VTm[\Psi][A\{\sigma\}]}}
\end{align*}
We can define the corresponding semantic term $M$ as follows:
\begin{align*}
  M &: \Sigma (\Psi' : \D) (\Sigma (\Sigma (\Sigma (\Sigma (\Sigma (s : \{ *\}) (\Psi : \D)) (\Phi :
      \D)) \\
    &\ \ (T : \STy(\Phi))) (t : \STm[\Phi][T])) (\sigma : \THom(\Psi, \Phi))) \\
  &\ \ \to \STm(\Psi, T\{\sigma\}) \\
  M(\_, (* , \Psi, \Phi, T, t, \sigma)) &:= t\{\sigma\}
\end{align*}
The substitution on the right hand side is given by the CwF structure in $\D$. 

\subsection{Internal Context Comprehension}

After finish internalizing all necessary components, we will show how to internalize
other operators. First let us consider context comprehension. Syntactically, we look
at internalizing the following judgment:
\begin{mathpar}
  \inferrule*
  { }
  {\typing[\Psi : \Ctx, T : \VTy(\Psi)]{\Psi.T}{\Ctx}}
\end{mathpar}
That is, we need to find a term to represent $\Psi.T$ when $\Psi$ and $T$ are
given:
\begin{align*}
  M &: \Sigma (\Phi : \D) (\Sigma(\Sigma (s : \{*\}) (\Psi : \D)) \STy(\Psi)) \to \D \\
  M(\_, (*, \Psi, T)) &:= \Psi.T
\end{align*}
where the comprehension on the right hand side is given by the CwF structure of
$\D$. This gives us the internal context comprehension.

\subsection{Internal Projection}

Next we show the existence of an internalized projection morphism. We consider a term
corresponding to the following judgment:
\begin{mathpar}
  \inferrule*
  { }
  {\typing[\Psi : \Ctx, T : \VTy[\Psi]]{p_T}{\UHom(\Psi.T, \Psi)}}
\end{mathpar}
We define the following term $p$:
\begin{align*}
  p &: \Sigma (\Phi : \D) (\Sigma(\Sigma (s : \{*\}) (\Psi : \D)) \STy(\Psi))
      \to \THom(\Psi.A, \Psi) \\
  p(\_, (*, \Psi, T)) &:= p_T
\end{align*}
$p_T$ on the right hand side is given by the CwF of $\D$ as before. 

\subsection{Internal Variables}

We consider the internalized version of variable projection:
\begin{mathpar}
  \inferrule*
  { }
  {\typing[\Psi : \Ctx, T : \VTy[\Psi]]{v_T}{\VTm[\Psi.T][T\{p_T\}]}}
\end{mathpar}
We define the following term $v$:
\begin{align*}
  v &: \Sigma(\Phi : \D)(\Sigma(\Sigma (s : \{*\}) (\Psi : \D)) \STy(\Psi)) \to
      \STm[\Gamma.T][T\{p_T\}] \\
  v(\_, (*, \Psi, T)) &:= v_T
\end{align*}
That is, $v$ is defined in terms of $v_T$ of $\D$.

\subsection{Internal Substitution Extension}

Next we consider internalized substitution extensions:
\begin{mathpar}
  \inferrule*
  { }
  {\typing[\Phi : \Ctx, \Psi : \Ctx, T : \VTy[\Psi], \sigma : \UHom(\Phi, \Psi), t :
    \VTm[\Phi][T\{\sigma\}]]{\ctxext{\sigma, t}}{\UHom(\Phi, \Psi.T)}}
\end{mathpar}

We define the following term:
\begin{align*}
  M &: \Sigma (\Phi' : \D) (\Sigma (\Sigma (\Sigma (\Sigma (\Sigma (s : \{ *\}) (\Phi : \D)) (\Psi :
      \D)) \\
    &\ \ (T : \STy(\Phi))) (\sigma : \THom(\Psi, \Phi))) (t : \STm[\Phi][T\{\sigma\}])) \\
    &\ \ \to \THom(\Phi, \Psi.T) \\
  M(\_, (*, \Phi, \Psi, T, \sigma, t))
    &:= \ctxext{\sigma, t}
\end{align*}
Here the right hand side is given by $\D$. 

\subsection{Programming $\D$ in $\whhat\D$}\label{sec:internal:q}

Previously, we have given the internalized version of the CwF structure of $\D$ in as
types and terms in $\whhat\D$. A natural consequence of the construction above is that
we can reason about $\D$ within $\whhat\D$ completely internally and that the result
of reasoning is totally faithful.

Consider a special morphism in any CwF structure. Given $\sigma : \UHom(\Phi, \Psi)$ and
$T : \VTy[\Psi]$, we can define a morphism $q(\sigma, T)$ by
\begin{align*}
  q(\sigma, T) &: \UHom(\Phi.T\{\sigma\}, \Phi.T) \\
  q(\sigma, T) &:= \ctxext{\sigma \circ p_{T\{\sigma\}}, v_{T\{\sigma\}}}
\end{align*}

If we assume that this definition is given in a syntactic type theory of $\whhat\D$
which complies with the judgments and the interpretations listed above, then by interpreting
this definition, we effectively obtain the following semantic term $M$ after expanding all
definitions:
\begin{align*}
  M &: \Sigma (\Phi' : \D) (\Sigma (\Sigma (\Sigma (\Sigma (s : \{ *\}) (\Phi : \D)) (\Psi :
      \D)) \\
    &\ \ (\sigma : \THom(\Phi, \Psi))) (T : \STy(\Psi))) \to \THom(\Phi.T\{\sigma\}, \Psi.T)  \\
  M(\_, (*, \Phi, \Psi, \sigma, T)) &= \ctxext{\sigma \circ p_{T\{\sigma\}}, v_{T\{\sigma\}}}
\end{align*}
Here the right hand side lives in $\D$. That is, the definition in a syntactic form of $\whhat\D$ is
reflected as is in the semantics in $\D$. 

\section{$\Pi$ Types in Presheaves}

In this section, let us consider another typical type former, $\Pi$ types. As
indicated by its name, $\Pi$ types model dependent function space of dependent type
theory. Again, each presheaf category has $\Pi$ types and we will review this fact in
this section.

\subsection{Definition}\label{sec:pi:def}

$\Pi$ types are an additional structure over a CwF.
\begin{definition}
  A CwF $\D$ has $\Pi$ types if it has the following data:
  \begin{enumerate}
  \item Given $A : \STy[\Gamma]$ and $\STy[\Gamma.A]$, $\Pi(A, B) : \STy[\Gamma]$,
  \item Given $M : \STm[\Gamma.A][B]$, $\Lambda(M) : \STm[\Gamma][\Pi(A, B)]$, 
  \item Given $M : \STm[\Gamma][\Pi(A, B)]$ and $N : \STm[\Gamma][A]$, $\TApp(M, N) :
    \STm[\Gamma][B\{id_\Gamma, N\}]$, 
  \end{enumerate}
  such that given $\sigma : \Delta \To \Gamma$, the following laws hold:
  \begin{enumerate}
  \item $\Pi(A, B)\{\sigma\} = \Pi(A\{\sigma\}, B\{q(\sigma, A)\})$,
  \item Given $M : \STm[\Gamma.A][B]$, $\Lambda(M)\{\sigma\} = \Lambda(M\{q(\sigma,
    A)\})$,
  \item Given $M : \STm[\Gamma][\Pi(A, B)]$ and $N : \STm[\Gamma][A]$, $\TApp(M,
    N)\{\sigma\} = \TApp(M\{\sigma\}, N\{\sigma\})$,
  \item Given $M : \STm[\Gamma.A][B]$ and $N : \STm[\Gamma][A]$, $\TApp(\Lambda(M), N)
    = M\{\ctxext{id_\Gamma, N}\}$.
  \end{enumerate}
\end{definition}

This definition models the invariance of types and terms of $\Pi$ under substitutions
as well as $\beta$ equivalence.

\subsection{An Incorrect Construction in Presheaves}

Given the definition of $\Pi$ types in a CwF, we are interested in revising the
constructions in \Cref{sec:presh}. Before considering a correct construction, let us
first check an incorrect construction and understand why we need a slightly more
complex definition.

We can get a first intuition by looking at how to derive from $M : \hSTm[\Gamma.A][B]$
$\whhat\Lambda(M) : \hSTm[\Gamma][\whhat\Pi(A, B)]$. We put hats on top of $\Pi$ and
$\Lambda$ to
indicate that the $\Pi$ types are constructed in $\whhat\D$.
\begin{align*}
  \hSTm[\Gamma.A][B] = \{ & M : (o : \Sigma (\Psi : \D) (\Sigma (s : \Gamma(\Psi))
                       A(\Psi, s))) \to B(o) \\
  | &\ \forall \delta : \Phi \To \Psi, s : \Gamma(\Psi), a : A(\Psi, s). \\
  & B(\delta, M(\Psi, (s, a))) = M(\Phi, (\Gamma(\delta, s), A(\delta, a))) \}
\end{align*}
For an overly simple-minded assumption, if $\whhat\Lambda(M)$ corresponds to currying,
it might be the following function:
\begin{align*}
  (o : \Sigma (\Psi : \D) (s : \Gamma(\Psi))) \to ((a : A(\Psi, s)) \to B(\Psi, s, a))
\end{align*}
This suggests the following formulation of $\whhat\Pi$ types:
\begin{align*}
  \whhat\Pi(A, B) &: (\smallint\Gamma)^{op} \To \Set \\
  \whhat\Pi(A, B)(\Psi, s) &:= (a : A(\Psi, s)) \to B(\Psi, s, a)
\end{align*}
However, this formulation is problematic because $(\Psi, s)$ occurs both in covariant and
contravariant position, and therefore we are not able to define the morphism part of
this functor. The solution to this problem turns out to be very similar to Kripke model
constructions of programming languages: if we view $\D$ as some sort of Kripke
structure, then we can generalize the object part to a function space which is
coherent under this Kripke relation, which is shown in the next subsection. 

\subsection{A Correct Construction}

\subsubsection{Definition of $\whhat\Pi$}

As shown in the previous subsection, the definition of $\whhat\Pi(A, B)$ must be
covariant in both $A$ and $B$. This can be achieved by defining a function space which
has coherence property w.r.t. $\D$-morphisms.
\begin{alignat*}{3}
  \whhat\Pi(A, B) &\span\omit$: (\smallint\Gamma)^{op} \To \Set$\hidewidth \\
  \whhat\Pi(A, B)(\Psi, s) &:= \{&& f : (\Phi : \D)(\delta : \D^{op}(\Psi, \Phi))(a : A(\Phi,
                                     \Gamma(\delta, s))) \to B(\Phi, (\Gamma(\delta, s),
                                     a)) \\
                                     & &&\ |\ P(A, B, \Psi, s, f) \}
\end{alignat*}
Here $P$ is a predicate over the function space, ensuring coherence of $f$. This
coherence condition is required in demonstrating right to left effect of the
isomorphism $\hSTm[\Gamma.A][B] \simeq \hSTm[\Gamma][\whhat\Pi(A, B)]$
below. Intuitively, it means $f$ is ``natural'': if the argument moves along the
Kripke relation, then the result also moves along the Kripke relation. 
\begin{align*}
  P(A, B, \Psi, s, f) :=&\ \forall \Phi : \D, \delta : \D^{op}(\Psi, \Phi), a : A(\Phi, \Gamma(\delta,
           s)), \Phi' : \D, \delta' : \D^{op}(\Phi, \Phi').  \\
  &\ f(\Phi', \delta' \circ \delta, A(\delta', a)) = B(\delta', f(\Phi, \delta, a))
\end{align*}
We examine the well-definedness of this predicate. $\delta'$ induces a morphism in
$(\smallint\Gamma)^{op}$: $(\smallint\Gamma)^{op}((\Phi, \Gamma(\delta, s)), (\Phi',
\Gamma(\delta' \circ \delta, s)))$. Therefore this gives
\begin{align*}
  A(\delta', a) &: A(\Phi', \Gamma(\delta' \circ \delta, s)) \\
  f(\Phi', \delta' \circ \delta, A(\delta', a)) &: B(\Phi', \Gamma(\delta' \circ
                                                  \delta, s), A(\delta', a))
\end{align*}
On the right hand side, $\delta'$ induces a morphism in $(\smallint(\Gamma.A))^{op}$,
from which we have the following analysis:
\begin{align*}
  \delta' &: (\smallint(\Gamma.A))^{op}((\Phi, \Gamma(\delta, s), a), (\Phi', \Gamma(\delta',
            \Gamma(\delta, s)), A(\delta', a))) \\
  f(\Phi, \delta, a) &: B(\Phi, (\Gamma(\delta, s), a)) \\
  B(\delta', f(\Phi, \delta, a)) &: B(\Phi', \Gamma(\delta', \Gamma(\delta, s)), A(\delta', a))
\end{align*}
Thus $P$ is well-defined.

Now we define the morphism part:
\begin{align*}
  \whhat\Pi(A, B)&(\delta : (\smallint\Gamma)^{op}((\Psi, s), (\Psi', \Gamma(\delta, s)))) \\
                 &(f : (\Phi : \D)(\delta : \D^{op}(\Psi, \Phi))(a : A(\Phi,
                   \Gamma(\delta, s))) \to B(\Phi, (\Gamma(\delta, s), a))) \\
                 &(\Phi, \delta' : \D^{op}(\Psi', \Phi), a : A(\Phi, \Gamma(\delta', \Gamma(\delta, s)))) \\
  :=&\ f(\Phi, \delta' \circ \delta, a) : B(\Phi, (\Gamma(\delta', \Gamma(\delta, s)), a))
\end{align*}
Notice that $\Gamma(\delta' \circ \delta, s) = \Gamma(\delta', \Gamma(\delta, s))$ due
to compositionality of $\Gamma$. Since $\whhat\Pi(A, B)(\delta, f)$ is defined in
terms of $f$, $P(A, B, \Psi', \Gamma(\delta, s), \whhat\Pi(A, B)(\delta, f))$ is inherited
from $P(A, B, \Psi, s, f)$.  Thus the morphism part of $\whhat\Pi(A, B)$ is well-defined.

\subsubsection{Definition of $\whhat\Lambda$}

Now let us define $\whhat\Lambda(M)$ for $M : \hSTm[\Gamma.A][B]$, the expansion of
which is given in the last subsection.  Now we define $\whhat\Lambda(M) :
\hSTm[\Gamma][\whhat\Pi(A, B)]$:
\begin{align*}
  \whhat\Lambda(M) &: \hSTm[\Gamma][\whhat\Pi(A, B)] \\
  \whhat\Lambda(M)(\Psi, s)(\Phi, \delta : \D^{op}(\Psi, \Phi), a : A(\Phi,
  \Gamma(\delta, s))) &:= M(\Phi, (\Gamma(\delta, s), a))
\end{align*}
Let us verify $P(A, B, \Psi, s, \whhat\Lambda(M)(\Psi, s))$. Given $\Phi', \delta' : \D^{op}(\Phi,
\Phi')$, we should prove
\begin{align*}
  M(\Phi', (\Gamma(\delta' \circ \delta, s), A(\delta', a))) = B(\delta', M(\Phi,
  (\Gamma(\delta, s), a)))
\end{align*}
This equation is directly proven by the specification of $M$. 

Let us verify the set specification for $\whhat\Lambda(M)$. It is required that the following equation holds
given any $\delta' : \Psi' \To \Psi, s : \Gamma(\Psi)$:
\begin{align*}
  \whhat\Pi(A, B)(\delta', \whhat\Lambda(M)(\Psi, s)) = \whhat\Lambda(M)(\Psi', \Gamma(\delta', s))
\end{align*}
We fully apply both sides:
\begin{align*}
  \whhat\Pi(A, B)(\delta', \whhat\Lambda(M)(\Psi, s))(\Phi, \delta, a) = \whhat\Lambda(M)(\Psi', \Gamma(\delta', s))(\Phi, \delta, a)
\end{align*}
We analyze the left hand side:
\begin{align*}
  &\whhat\Pi(A, B)(\delta', \whhat\Lambda(M)(\Psi, s))(\Phi, \delta, a) \\
  =&\ \whhat\Lambda(M)(\Psi, s)(\Phi, \delta \circ \delta', a) \tag*{by morphism part of $\whhat\Pi(A, B)$}
  \\
  =&\ M(\Phi, (\Gamma(\delta \circ \delta', s), a)) \tag*{by definition of $\whhat\Lambda(M)$}
\end{align*}
Then the right hand side:
\begin{align*}
  & \whhat\Lambda(M)(\Psi', \Gamma(\delta', s))(\Phi, \delta, a) \\
  =&\ M(\Phi, (\Gamma(\delta, \Gamma(\delta', s)), a))
     \tag*{by definition of $\whhat\Lambda(M)$} \\
  =& M(\Phi, (\Gamma(\delta \circ \delta', s), a)) \tag*{by functoriality of $\Gamma$}
\end{align*}
We have shown that $\whhat\Lambda(M)$ does reside in set $\hSTm[\Gamma][\whhat\Pi(A, B)]$. 

\subsubsection{$\hSTm[\Gamma.A][B] \simeq \hSTm[\Gamma][\whhat\Pi(A, B)]$}

Next, we show that $\hSTm[\Gamma.A][B]$ and $\hSTm[\Gamma][\whhat\Pi(A, B)]$ are
isomorphic. We have obtained the left to right directly via $\whhat\Lambda$. Now let
us consider the other direction $\whhat{\Lambda^{-1}}$.

Assuming $M' : \hSTm[\Gamma][\whhat\Pi(A, B)]$, we have
\begin{align*}
  \whhat{\Lambda^{-1}}(M') &: \hSTm[\Gamma.A][B] \\
  \whhat{\Lambda^{-1}}(M')(\Psi, (s, a)) &:= M'(\Psi, s)(\Psi, id_\Psi, a)
\end{align*}
The specification requires that given any $\delta : \D^{op}(\Psi, \Psi'), s : \Gamma(\Psi),
a : A(\Psi, s)$:
\begin{align*}
  B(\delta, \whhat{\Lambda^{-1}}(M')(\Psi, (s, a))) = \whhat{\Lambda^{-1}}(M')(\Psi', \Gamma.A(\delta, (s, a)))
\end{align*}
We expand this equation and get:
\begin{align*}
  B(\delta, M'(\Psi, s)(\Psi, id_\Psi, a)) = M'(\Psi', \Gamma(\delta, s))(\Psi', id_{\Psi'}, A(\delta, a))
\end{align*}
We analyze the left hand side:
\begin{align*}
  &\ B(\delta, M'(\Psi, s)(\Psi, id_\Psi, a)) \\
  =&\ M'(\Psi, s)(\Psi', \delta \circ id_\Psi, A(\delta, a))
     \tag*{$M'(\Psi, s)$ satisfies $P$} \\
  =&\ M'(\Psi', \Gamma(\delta, s))(\Psi', id_{\Psi'}, A(\delta, a))
     \tag*{by specification of $M'$}
\end{align*}
This concludes $\whhat{\Lambda^{-1}}(M')$ is well defined. 

Now we proceed to proving that $\whhat\Lambda$ and $\whhat{\Lambda^{-1}}$ do form the
effects of the isomorphism $\hSTm[\Gamma.A][B] \simeq \hSTm[\Gamma][\whhat\Pi(A, B)]$.

Given $M : \hSTm[\Gamma.A][B]$,
\begin{align*}
  \whhat{\Lambda^{-1}}(\whhat\Lambda(M))(\Psi, (s, a))
  &= \whhat\Lambda(M)(\Psi, s)(\Psi, id_\Psi, a)
    \tag*{by definition of $\whhat{\Lambda^{-1}}$} \\
  &= M(\Psi, (s, a)) \tag*{by definition of $\whhat\Lambda$}
\end{align*}

Given $M' : \hSTm[\Gamma][\whhat\Pi(A, B)]$,
\begin{align*}
  \whhat\Lambda(\whhat{\Lambda^{-1}}(M'))(\Psi, s)(\Phi, \delta, a)
  &= \whhat{\Lambda^{-1}}(M')(\Phi, (\Gamma(\delta, s), a))
    \tag*{by definition of $\whhat\Lambda$} \\
  &= M'(\Phi, \Gamma(\delta, s))(\Phi, id_\Phi, a)
    \tag*{by definition of $\whhat{\Lambda^{-1}}$} \\
  &= M'(\Psi, s)(\Phi, \delta, a)
    \tag*{by specification of $M'$}
\end{align*}
We conclude the intended isomorphism.

\subsubsection{Definition of $\whhat{\TApp}$}

Finally we define the semantic applications. Given
$M : \hSTm[\Gamma][\whhat\Pi(A, B)]$ and $N : \hSTm[\Gamma][A]$ we define
\begin{align*}
  \whhat{\TApp}(M, N) &: \hSTm[\Gamma][B\{\ctxext{id_\Gamma, N}\}] \\
  \whhat{\TApp}(M, N) &:= \whhat{\Lambda^{-1}}(M)\{\ctxext{id_\Gamma, N}\}
\end{align*}

\subsection{Laws}

We have got all the definitions. The next step is to verify that the definitions
satisfy the laws. We verify the laws in the order as given in \Cref{sec:pi:def}.
We assume a substitution morphism $\sigma : \Delta \To \Gamma$ when necessary.

\begin{enumerate}
\item We shall show the following equation of substitution of $\whhat\Pi$ types:
  \begin{align*}
    \whhat\Pi(A, B)\{\sigma\} = \whhat\Pi(A\{\sigma\}, B\{q(\sigma, A)\})
  \end{align*}
  We expand the left hand side:
  \begin{align*}
    &\ \whhat\Pi(A, B)\{\sigma\}(\Psi, s) \\
    =&\ \whhat\Pi(A, B)(\Psi, \sigma(\Psi, s))
      \tag*{by definition of type substitutions} \\
    =&\ \{f : (\Phi : \D)(\delta : \D^{op}(\Psi, \Phi))(a : A(\Phi,
       \Gamma(\delta, \sigma(\Psi, s)))) \\
     &\ \ \to B(\Phi, (\Gamma(\delta, \sigma(\Psi, s)), a))\ |\ P(A, B, \Psi, \sigma(\Psi, s),
       f) \}
  \end{align*}
  By naturality of $\sigma$, we have $\Gamma(\delta, \sigma(\Psi, s)) = \sigma(\Phi, \Delta(\delta, s))$. Thus
  \begin{align*}
    &\ \whhat\Pi(A, B)\{\sigma\}(\Psi, s) \\
    =&\ \{f : (\Phi : \D)(\delta : \D^{op}(\Psi, \Phi))(a : A(\Phi,
       \sigma(\Phi, \Delta(\delta, s)))) \\
     &\ \ \to B(\Phi, (\sigma(\Phi, \Delta(\delta, s)), a))\ |\ P(A, B, \Psi, \sigma(\Psi, s),
       f) \} \\
    =&\ \{f : (\Phi : \D)(\delta : \D^{op}(\Psi, \Phi))(a : A\{\sigma\}(\Phi, \Delta(\delta, s))) \\
     &\ \ \to B\{q(\sigma, A)\}(\Phi, (\Delta(\delta, s), a))\ |\ P(A, B, \Psi, \sigma(\Psi, s),
       f) \} 
  \end{align*}
  The function signature matches up. We just need to show that \linebreak
  $P(A, B, \Psi, \sigma(\Psi, s), f)$ and
  $P(A\{\sigma\}, B\{q(\sigma, A)\}, \Psi, s, f)$ are equivalent. We can see that
  these two propositions are essentially the same. This is because
  \begin{enumerate}
  \item Naturality of $\sigma$ leads to
    $A(\Phi, \Gamma(\delta, \sigma(\Psi, s))) = A(\Phi, \sigma(\Phi, \Delta(\delta,
    s)))$ and thus the arguments of $a$ in both propositions are the same;
  \item $B\{\sigma\}(\delta') = B(\delta')$ by definition.
  \end{enumerate}
  Now we conclude the target equation. 
\item For $M : \hSTm[\Gamma.A][B]$, we should prove
  $\whhat\Lambda(M)\{\sigma\} = \whhat\Lambda(M\{q(\sigma, A)\})$. We fully apply and
  expand both sides:
  \begin{align*}
    &\ \whhat\Lambda(M)\{\sigma\}(\Psi, s)(\Phi, \delta, a) \\
    =&\ \whhat\Lambda(M)(\Psi, \sigma(\Psi, s))(\Phi, \delta, a)
       \tag*{by definition of term substitutions} \\
    =&\ M(\Phi, (\Gamma(\delta, \sigma(\Psi, s)), a))
       \tag*{by definition of $\whhat\Lambda$}
  \end{align*}
  \begin{align*}
    &\ \whhat\Lambda(M\{q(\sigma, A)\})(\Psi, s)(\Phi, \delta, a) \\
    =&\ M\{q(\sigma, A)\}(\Phi, (\Gamma(\delta, s), a))
       \tag*{by definition of $\whhat\Lambda$} \\
    =&\ M(\Phi, (\Gamma(\delta, \sigma(\Psi, s)), a))
       \tag*{by definition of term substitutions}
  \end{align*}
  Thus the target equation holds. 
\item For $M : \STm[\Gamma][\Pi(A, B)]$ and $N : \STm[\Gamma][A]$,
  $\TApp(M, N)\{\sigma\} = \TApp(M\{\sigma\}, N\{\sigma\})$. We extend both sides:
  \begin{align*}
    &\ \TApp(M, N)\{\sigma\} \\
    =&\ \whhat{\Lambda^{-1}}(M)\{\ctxext{id_\Gamma, N}\}\{\sigma\}
       \tag*{by definition of $\TApp$} \\
    =&\ \whhat{\Lambda^{-1}}(M)\{\ctxext{\sigma, N\{\sigma\}}\}
       \tag*{property of substitution extension} 
  \end{align*}
  \begin{align*}
    \TApp(M\{\sigma\}, N\{\sigma\}) =
    \whhat{\Lambda^{-1}}(M\{\sigma\})\{\ctxext{id_\Delta, N\{\sigma\}}\}
    \tag*{by definition of $\TApp$}
  \end{align*}

  We prove a more general statement in order to conclude the target equation. Given
  $\sigma' : \Delta \To \Gamma$ and $N' : \hSTm[\Delta][A\{\sigma\}]$, the following
  equation holds:
  \begin{align*}
    \whhat{\Lambda^{-1}}(M)\{\ctxext{\sigma, N'}\} =
    \whhat{\Lambda^{-1}}(M\{\sigma\})\{\ctxext{id_\Delta, N'}\}
  \end{align*}
  We fully apply and expand both sides:
  \begin{align*}
    &\ \whhat{\Lambda^{-1}}(M)\{\ctxext{\sigma, N'}\}(\Psi, (s, a)) \\
    =&\ \whhat{\Lambda^{-1}}(M)(\Psi, (\sigma(\Psi, s), N'(\Psi, a)))
       \tag*{by term substitution and substitution extension} \\
    =&\ M(\Psi, \sigma(\Psi, s))(\Psi, id_\Psi, N'(\Psi, a))
       \tag*{by definition of $\whhat{\Lambda^{-1}}$}
  \end{align*}
  \begin{align*}
    &\ \whhat{\Lambda^{-1}}(M\{\sigma\})\{\ctxext{id_\Delta, N'}\}(\Psi, (s, a)) \\
    =&\ \whhat{\Lambda^{-1}}(M\{\sigma\})(\Psi, (s, N'(\Psi, a)))
       \tag*{by term substitution and substitution extension} \\
    =&\ M\{\sigma\}(\Psi, s)(\Psi, id_\Psi, N'(\Psi, a))
       \tag*{by definition of $\whhat{\Lambda^{-1}}$} \\
    =&\ M(\Psi, \sigma(\Psi, s))(\Psi, id_\Psi, N'(\Psi, a))
       \tag*{by term substitution}
  \end{align*}
  That concludes the statement and thus the target equation. 
  
\item $\whhat{\TApp}(\whhat\Lambda(M), N) = M\{\ctxext{id, N}\}$ holds automatically
  due to the isomorphism $\hSTm[\Gamma.A][B] \simeq \hSTm[\Gamma][\whhat\Pi(A, B)]$.
\end{enumerate}

\subsection{Internal $\Pi$ Types}

In this section we consider internalizing $\Pi$ types. In the previous section, we
showed that all presheaf categories have CwF structures equipped with $\whhat\Pi$
types. Continuing the discussion in \Cref{sec:internal}, by assuming $\Pi$ types in
$\D$, we can also internalize $\Pi$ types and related terms in $\whhat\D$ as well.
Internalization implies providing interpretation for the following judgments:
\begin{mathpar}
  \inferrule*
  { }
  {\typing[\Psi : \Ctx, S : \VTy[\Psi], T : \VTy[\Psi.S]]{\Pi(S, T)}{\VTy[\Psi]}}

  \inferrule*
  { }
  {\typing[\Psi : \Ctx, S : \VTy[\Psi], T : \VTy[\Psi.S], t :
    \VTm[\Psi.S][T]]{\Lambda(t)}{\VTm[\Psi][\Pi(S, T)]}}

  \inferrule*
  { }
  {\typing[\Psi : \Ctx, S : \VTy[\Psi], T : \VTy[\Psi.S], t :
    \VTm[\Psi][\Pi(S, T)], s : \VTm[\Psi][S]]{\UApp(t,
      s)}{\VTm[\Psi][T\{\ctxext{id_\Psi, s}\}]}}
\end{mathpar}

All these judgments correspond to semantic terms in $\whhat\D$ of the matching types.
We define the terms in the style given in \Cref{sec:internal}. Let us first define the
term $M$ for $\Pi$:
\begin{align*}
  M &: \Sigma(\Psi' : \D) (\Sigma (\Sigma (\Sigma (\_ : \{*\}) (\Psi : \D)) \\
    &\ \ (S : \STy[\Psi])) (T : \STy[\Psi.S])) \to \STy[\Psi] \\
  M(\_, (*, \Psi, S, T)) &:= \Pi(S, T)
\end{align*}

Next we give the term $M'$ for the internal $\Lambda$:
\begin{align*}
  M' &: \Sigma(\Psi' : \D) (\Sigma (\Sigma (\Sigma (\Sigma (\_ : \{*\}) (\Psi : \D)) \\
    &\ \ (S : \STy[\Psi])) (T : \STy[\Psi.S])) (t : \STm[\Psi.S][T])) \to
      \STm[\Psi][\Pi(S, T)] \\
  M'(\_, (*, \Psi, S, T, t)) &:= \Lambda(t)
\end{align*}

Next we give the term $M''$ for the internal $\UApp$:
\begin{align*}
  M'' &: \Sigma(\Psi' : \D) (\Sigma (\Sigma (\Sigma (\Sigma (\Sigma (\_ : \{*\}) (\Psi
        : \D)) (S : \STy[\Psi])) \\
      &\ \ (T : \STy[\Psi.S])) (t : \STm[\Psi][\Pi(S, T)])) (s : \STm[\Psi][S])) \\
      &\ \ \to \STm[\Psi][T\{\ctxext{id_\Psi, s}\}] \\
  M''(\_, (*, \Psi, S, T, t, s)) &:= \UApp(t, s)
\end{align*}

$\Pi$, $\Lambda$ and $\UApp$ on the right are given by the $\Pi$ structure of $\D$. 
Thus the $\Pi$ structure of $\D$ can also be internalized. 

\section{Internalize Closed Types and Terms}

In \Cref{sec:internal}, we showed that the presheaf category can internalize $\D$ and
its CwF structure. These internal types and terms are relative to an internal
context in the base category and thus these internal types and terms can be potentially
open. To obtain closed types and terms, we can set the internal context to be the
empty one:
\begin{mathpar}
  \inferrule*
  { }
  {\istype[]{\VTy[\top]}}

  \inferrule*
  { }
  {\istype[T : \VTy[\top]]{\VTm[\top][T]}}
\end{mathpar}

\subsection{Closed Types}

There is another way to define semantic types to represent closed internal types and terms. Notice that in \Cref{sec:internal}, all the semantic types are defined by
ignoring the first projection. If we view the base category $\D$ as an index category
for a Kripke relation, then that means the semantic types we defined in
\Cref{sec:internal} are invariant under the Kripke relation. This allows us to
faithfully internalize all structures of $\D$ in $\whhat\D$. Hence, it is interesting to
consider what if a semantic type respects the Kripke relation?  This motivates a
different way to internalize types and terms in $\D$ in $\whhat\D$. 
\begin{align*}
  \VTy' &: \hSTy[\wtop] \\
  \VTy'(\Psi, *) &:= \STy(\Psi) \\
  \VTy'(\delta : (\smallint \wtop)^{op}((\Psi, *), (\Phi, *))) &:= S
                                                                 \mapsto S\{\delta\} :
                                                                 \STy(\Psi) \to \STy[\Phi]
\end{align*}

This semantic type interprets the following judgment:
\begin{mathpar}
  \inferrule*
  { }
  {\istype[]{\VTy'}}
\end{mathpar}
Now let us examine what constitutes its semantic terms. By definition,
\begin{align*}
  \hSTm[\wtop][\VTy'] = \{& M : \Sigma (\Psi : \D) \{ * \} \to \STy(\Psi) \\
                          &|\  \forall
                            \Phi : \D, \delta : \D^{op}(\Psi, \Phi). M(\Psi, *)\{\delta\}
                            = M(\Phi, *) \}
\end{align*}
That is, if $M : \hSTm[\wtop][\VTy']$, then $M(\Psi, *)$ is a valid type in $\D$ for
all $\Psi$ and is coherent with substitution in $\D$. Effectively, this implies $M$
represents a \emph{closed} type in $\D$, because once $M(\top, *)$ is fixed,
$M(\Psi, *)$ for any $\Psi$ is induced due to uniqueness of $\Psi \To \top$. Thus we have
\begin{align*}
  \hSTm[\wtop][\VTy'] \simeq \STy(\top) \simeq \hSTm[\wtop][\VTy[\top]]
\end{align*}
That is $\hSTm[\wtop][\VTy']$ is isomorphic to $\hSTm[\wtop][\VTy[\top]]$. 

\subsection{Closed Terms}

We can construct closed internal terms in a very similar way. The corresponding
syntactic type is
\begin{mathpar}
  \inferrule*
  { }
  {\istype[T : \VTy']{\VTm'(T)}}
\end{mathpar}
This semantic type is constructed as follows:
\begin{align*}
  \VTm' &: \hSTy[\wtop.\VTy'] \\
  \VTm'(\Psi, (*, T)) &:= \STm(\Psi, T) \\
  \VTm'(\delta : (\smallint (\wtop.\VTy'))^{op}((\Psi, (*, T)), (\Phi, (*,
  T\{\delta\})))) &:= s : \STm(\Psi, T) \mapsto s\{\delta\} : \STm[\Psi][T\{\delta\}]
\end{align*}
Similar to $\VTy'$, the morphism part of the functorial action of $\VTm'$ is
substitutions. Let us consider the semantic terms of $\VTm'$:
\begin{align*}
  \hSTm[\wtop.\VTy'][\VTm'] = \{& M : \Sigma (\Psi : \D) (\Sigma \{*\} (T :
                                  \STy(\Psi))) \to \STm(\Psi, T) \\
                                &|\ \forall \Phi : \D, \delta : \D^{op}(\Psi, \Phi), T : \STy(\Psi). \\
  &\ \ M(\Psi, (*,
    T))\{\delta\} = M(\Phi, (*, T\{\delta\}))  \}
\end{align*}
We are not able to show that $\hSTm[\wtop.\VTy'][\VTm']$ corresponds to closed terms
from the specifications directly, because we do not know $T$ is a (weakened) closed
type. Nevertheless, we can be sure about this because the definition of
$\hSTm[\wtop][\VTy']$ implicitly requires $T$ to be closed, as previously
discussed.

\subsection{Closed versus Open}

Now we see that $\VTy$ and $\VTm$ are strictly more expressive than $\VTy'$ and
$\VTm'$. In particular, $\VTy'$ and $\VTm'$ are closed, so there cannot be any
frontend syntax for any binding construct.  What can they be used for?  Since they are
closed, one kind of systems they can model is typed combinatorics. Since we do not
need variables in combinatorics, we only need to talk about closed programs. Moreover,
the operations for $\VTy'$ and $\VTm'$ are much limited. We are not able to express
internal substitutions and their operations because we have no direct access
to the underlying $\D$-morphisms. 

\section{Constant Presheaves}

All previous constructions are coherent with $\D$-morphisms due to functoriality or
set specifications. In this section, we consider a special class of functors which
have a ``world'' fixed:
\begin{align*}
  \square &: \whhat\D \To \whhat\D \\
  \square(\Gamma)(\Psi : \D) &:= \Gamma(\top) \\
  \square(\Gamma)(\delta, s) &:= s
\end{align*}
We refer to this functor as $\square$ because it is a comonadic modality. We can
easily see this by construction:
\begin{align*}
  \square\square(\Gamma) &= \square \Gamma \\
  \epsilon_\square &: \square(\Gamma) \To \Gamma \\
  \epsilon_\square(\Psi : \D, s : \Gamma(\top)) &:= \Gamma(!, s)
\end{align*}
$\square$ is a definitionally idempotent comonad, which is not the most general
necessity modality. In the following definitions, we make use of the idempotency, so
the construction is not generally applicable to other comonadic modality. A general
formulation is left for future investigation.

\subsection{Semantic Types and Terms}

Next we consider some constructions. Given a semantic type $A : \hSTy[\Gamma]$, we obtain
\begin{align*}
  \square A &: \hSTy[\square \Gamma] \\
  \square A(\Psi, s : \Gamma(\top)) &:= A(\top, s) \\
  \square A(\delta) &:= M \mapsto M
\end{align*}
When we apply context comprehension to $\square\Gamma$ and $\square A$, we have
\begin{align*}
  \square\Gamma.\square A(\Psi) &= \Sigma(s : \Gamma(\top))A(\top, s) \\
  \square\Gamma.\square A(\delta)(s, a) &= (s, a)
\end{align*}
Thus we have $\square \Gamma . \square A = \square (\Gamma.A)$. 

Given a semantic term $M : \hSTm[\Gamma][A]$, we have
\begin{align*}
  \square M &: \hSTm[\square \Gamma][\square A] \\
  \square M(\Psi, (s : \Gamma(\top), a : A(\top, s))) &:= M(\top, (s, a))
\end{align*}
Since the morphism action of $\square T$ is identity, the specification of $\square M$
is trivially true. 

\subsection{Interpreting Introduction Rule}

We can interpret the introduction rule for necessity:
\begin{mathpar}
  \inferrule*
  {\mtyping[\Delta][\cdot]{M}{A}}
  {\mtyping{\ubox M}{\square A}}
\end{mathpar}
We interpret the global context to $\square \Delta$ and $\Delta;\Gamma$ to
$\square \Delta; \Gamma$ in $\whhat\D$, where $;$ in the semantic means concatenation. This
operation is defined by recursively interpreting $\Gamma$:
\begin{align*}
  \Delta;\top &:= \Delta \\
  \Delta;(\Gamma.A) &:= (\Delta;\Gamma).A
\end{align*}
The operator $;$ has lower precedence than context comprehension. That means
$\Delta;\Gamma.A$ means $\Delta;(\Gamma.A)$. 
Given $M : \hSTm[\square\Delta][A]$ and $k = | \Gamma |$, we obtain
\begin{align*}
  \tbox(M) &: \hSTm[\square\Delta;\Gamma][\square A \{ p^k\}] \\
   \tbox(M) &:= \square M \{ p^k \}
\end{align*}
This formulation is special and takes advantage of the idempotency of $\square$,
because we implicitly make use of $\square \square \Delta = \square \Delta$.

\subsection{Interpreting Elimination Rule}

For the eliminator, we can explore two different flavors:
\begin{mathpar}
  \ismtype[\Delta][\cdot]{A}
  
  \inferrule*
  {\ismtype[\Delta][\Gamma, x : \square A]{B} \\ \mtyping{M}{\square A} \\
    \mtyping[\Delta, u : A]{N}{B[\ubox u/x]}}
  {\mtyping{\uletbox u M N}{B[M/x]}}

\end{mathpar}
The formulation is given by \cite{DBLP:journals/mscs/Shulman18}. 
Let us interpret the syntax to the model.

Given $A : \hSTy[\square \Delta]$,
$B : \hSTy[\square \Delta; \Gamma.\square A\{p^k\}]$ for $k = |\Gamma|$,
$M : \hSTm[\square \Delta;\Gamma][\square A\{p^k\}]$ and
$N : \hSTm[\square (\Delta.A);\Gamma\{p\}][B\{\ctxext{q(p, \Gamma), v_{\square
    A}\{p^k\}}\}]$, we want to construct
$\tletbox(M, N) : \hSTm[\square \Delta; \Gamma][B\{\ctxext{id_{\square \Delta;
    \Gamma}, M}\}]$. When stating the semantic term of $N$, we used certain
generalizations, which we make explicit below. $q(\sigma, \Gamma)$ is a generalization
of $q(\sigma, A)$ defined in \Cref{sec:internal:q}. $\Gamma\{\sigma\}$ is the
generalization of type substitution. Given $\sigma : \Delta' \To \Delta$ and that
$\Delta;\Gamma$ is a valid context, we define
$q(\sigma, \Gamma) : \Delta';\Gamma\{\sigma\} \To \Delta;\Gamma$ and
$\Delta';\Gamma\{\sigma\}$ mutually:
\begin{align*}
  q(\sigma, \top)  &:= \sigma \\
  q(\sigma, \Gamma.A)  &:= q(q(\sigma, \Gamma), A) \\
  \Delta';\top\{\sigma\} &:= \Delta' \\
  \Delta';(\Gamma.A)\{\sigma\} &:= \Delta';\Gamma\{\sigma\}.A\{q(\sigma, \Gamma)\}
\end{align*}
Next we can move on to defining $\tletbox(M, N)$:
\begin{align*}
  \tletbox(M, N) &: \hSTm[\square \Delta; \Gamma][B\{\ctxext{id_{\square \Delta;
                   \Gamma}, M}\}] \\
  \tletbox(M, N)(\Psi, (s : \Delta(\top), s' : \Gamma(\Psi)))
                 &:= N(\Psi, (s, M(\Psi, (s, s')) : A(\top, s), s'))
\end{align*}
Since we interpret global contexts as $\square$ contexts, $M$ of type $\square A$ can
be directly applied to $A$ in the global context $\square (\Delta.A)$. This allows us
to directly plug in $M(\Psi, (s, s'))$ in an argument of $N$. This gives us one model
for necessity. 


\section{Conclusion}

In this note, we reviewed a construction of CwF structure in any presheaf category using
category of elements. We showed that any syntactic type theory interpreting the CwF
structure of a presheaf category can be used to reason about the base
category. Moreover, the syntactic representation of the base category is faithful. We
also showed that this construction extends to $\Pi$ types. We also consider a
idempotent comonadic modality in a presheaf category. 

\bibliography{ref}

\begin{thebibliography}{}

\bibitem[Dybjer, 1995]{DBLP:conf/types/Dybjer95}
Dybjer, P. (1995).
\newblock Internal type theory.
\newblock In {\em Types for Proofs and Programs, International Workshop
  TYPES'95, Torino, Italy, June 5-8, 1995, Selected Papers}, pages 120--134.

\bibitem[Hofmann, 1997]{Hofmann1997}
Hofmann, M. (1997).
\newblock {\em Syntax and semantics of dependent types}, pages 13--54.
\newblock Springer London, London.

\bibitem[Shulman, 2018]{DBLP:journals/mscs/Shulman18}
Shulman, M. (2018).
\newblock Brouwer's fixed-point theorem in real-cohesive homotopy type theory.
\newblock {\em Math. Struct. Comput. Sci.}, 28(6):856--941.

\end{thebibliography}

\end{document}